\documentclass[reprint,aps,amsmath,amssymb,twoside,prd,showkeys,superscriptaddress]{revtex4-1}
\pdfoutput=1
\usepackage{verbatim}
\usepackage{comment}
\usepackage{graphicx}
\usepackage[T1]{fontenc}
\usepackage{epsfig}
\usepackage{bm}
\usepackage{tensor}
\usepackage{amssymb}
\usepackage{float}
\usepackage{amsmath}
\usepackage{subfigure}
\usepackage{dcolumn}
\usepackage[utf8]{inputenc}
\usepackage{cancel}
\usepackage[colorlinks]{hyperref}
\usepackage[usenames,dvipsnames]{color}
\hypersetup{
     breaklinks=true,
    pdfstartview={FitH},    
    colorlinks=true,       
    linkcolor=blue,          
    citecolor=red,        
    filecolor=magenta,      
    urlcolor=blue,           
    anchorcolor=green,      
    linktocpage=true
}

\def\doi{http://doi.org}

\newcommand{\HCd}{\mathcal{H}}

\def\HCdt0{\tilde{\HCd}_{0}}

\newcommand{\afffias}{Frankfurt Institute for Advanced Studies (FIAS), 
Ruth-Moufang-Strasse~1, 60438 Frankfurt am Main, Germany}
\newcommand{\affbgu}{Physics Department, Ben-Gurion University of the Negev, 
Beer-Sheva 
84105, Israel}
\newcommand{\affbahamas}{Bahamas Advanced Study Institute and Conferences, 4A 
Ocean 
Heights, Hill View Circle, Stella Maris, Long Island, The Bahamas}

\newcommand{\afflon}{School of Mathematical Sciences, Queen Mary University of London, Mile End Road, London, E1 4NS, United Kingdom}

\newcommand{\affbarc}{Departament de Matem\`atiques, Universitat Polit\`ecnica de Catalunya, Diagonal 647, 08028 Barcelona, Spain}
\newcommand{\affcam}{DAMTP, Centre for Mathematical Sciences, University of Cambridge, Wilberforce Road, Cambridge CB3 0WA, United Kingdom}
\begin{document}
\title{$\alpha$-attractors in Quintessential Inflation motivated by Supergravity}
\author{L. Arest\'e Sal\'o}
\email{l.arestesalo@qmul.ac.uk}
\affiliation{\afflon}
\author{D. Benisty}
\email{benidav@post.bgu.ac.il}
\affiliation{\affbgu}\affiliation{\afffias}\affiliation{\affcam}
\author{E. I. Guendelman}
\email{guendel@bgu.ac.il}
\affiliation{\affbgu}\affiliation{\afffias}\affiliation{\affbahamas}
\author{J. d. Haro}
\email{jaime.haro@upc.edu}
\affiliation{\affbarc}
\begin{abstract}
An exponential kind of quintessential inflation potential motivated by supergravity is studied. This type belongs to the class of $\alpha$-attractor models. The model was studied for the first time in \cite{Dimopoulos:2017zvq} in which the authors introduced a negative Cosmological constant in order to ensure a zero-vacuum energy density at late times. However, in this paper, we disregard this cosmological constant, showing that the obtained results are very close to the ones obtained recently in the context of Lorentzian quintessential inflation and thus depicting with great accuracy the early and late-time acceleration of our Universe. The model is compatible with the recent observations. Finally, we review the treatment of the $\alpha $ attractor and we show that our potential depicts the late time cosmic acceleration with an effective equation of state equal to $-1$.

\end{abstract}

\maketitle

\section{Introduction}
The inflationary paradigm is considered as a necessary part of the standard model of 
cosmology, since it provides the solution to the the horizon, the flatness, and the monopole problems
\cite{Guth:1980zm,Guth:1982ec,Starobinsky:1979ty,Kazanas:1980tx,Starobinsky:1980te,Linde:1981mu,Albrecht:1982wi,Barrow:1983rx,Blau:1986cw}. It can be achieved through various mechanisms, for instance through the introduction of a scalar inflaton field \cite{Barrow:2016qkh,Barrow:2016wiy,Olive:1989nu,Linde:1993cn,Liddle:1994dx,Germani:2010gm,Kobayashi:2010cm,Feng:2010ya,Burrage:2010cu,Kobayashi:2011nu,Ohashi:2012wf,Cai:2014uka,Kamali:2016frd,Benisty:2017lmt,Middleton:2019bio,Dalianis:2018frf,Dalianis:2019asr,Qiu:2020qsq,Choudhury:2014uxa,Choudhury:2013iaa,Choudhury:2011sq,Choudhury:2013jya,Choudhury:2014sxa,Choudhury:2017cos,Banerjee:2020xcn}. The well-known Starobinsky model \cite{starobinsky} (see also \cite{riotto1} for a review), originally conceived to find non-singular solutions going beyond General Relativity (GR) -although it really only contains one unstable non-singular solution- is one of the best candidates to depict correctly the inflationary period, in the sense that the theoretical results provided by the model match perfectly with the current observation data \cite{planck,planck18,planck18a}. An important extension of the Starobinsky model, coming from supergravity, are the so-called $\alpha$-attractors \cite{linde1,kallosh,kallosh1,kallosh2,kallosh3, Linde,ferrara, kallosh4,Canas-Herrera:2021sjs}, which also depict very well inflation, and are used for the first time in the context of Quintessential Inflation \cite{pv,dimopoulos1,hossain2,Geng, hossain1, hap19, hap1, deHaro, AresteSalo, Guendelman4} for a detailed explanation of the early and recent results of this topic -  in \cite{Dimopoulos:2017zvq} (see also \cite{Akrami:2017cir}). In that paper, the authors introduce a negative Cosmological Constant (CC) in order to have at late times an exponential potential which guarantees an eternal acceleration for a wide range of the parameters involved in the model or getting a transient acceleration at the present time. Here we see that for this simple model, which only depends on two parameters, quintessential inflation is also obtained without the introduction of this CC. In fact, the model leads to the same results provided by Lorentzian Quintessential Inflation \cite{Guendelman,Guendelman1, Guendelman2}, i.e., the model behaves as an $\alpha$-attractor at early times and provides, at late times, an eternal inflation with an effective Equation of State (EoS) parameter equal to $-1$ at very late times.

The paper is organized as follows: In Section \ref{sec-2} we calculate the Power Spectrum of perturbations and the value of the parameters involved in the model in agreement with the observational data at early and late times.
Section \ref{sec-3} is devoted to the analytic calculation of the value of the field and its derivative at the reheating time, which is needed to perform the numerical calculations up to the present and future times. In Section \ref{sec-4} we perform the numerical calculations, showing that at late times the Universe enters in an eternal acceleration, and, at the present time, the effective EoS parameter, given by the model, agrees with the Planck's results. The introduction of a CC is done in Section \ref{sec-5}, where we review the work done in \cite{Dimopoulos:2017zvq},  showing that the model with a CC is only compatible with the Planck's observational data for a narrow range of values of the parameter $\alpha$, which, from our point of view,  does not prove its viability. 

The units used throughout the paper are $\hbar=c=1$ and we denote  the reduced Planck's mass by 
$M_{pl}\equiv \frac{1}{\sqrt{8\pi G}}\cong 2.44\times 10^{18}$ GeV.

\section{$\alpha$-attractors in quintessential inflation}
\label{sec-2}

We consider the following Lagrangian motivated by supergravity and corresponding to a non-trivial K\"ahler manifold (see for instance \cite{Dimopoulos:2017zvq} and the  references therein), combined with  an standard   exponential potential,
\begin{eqnarray}\label{lagrangian}
\mathcal{L}=\frac{1}{2}\frac{\dot{\phi}^2}{(1-\frac{\phi^2}{6\alpha}  )^2}M_{pl}^2-\lambda M_{pl}^4 e^{-\kappa \phi},
\end{eqnarray}
where $\phi$ is a dimensionless scalar field, and $\kappa$ and $\lambda$ are positive dimensionless constants.

In order that the kinetic term has the canonical  form, one can redefine the scalar field as follows,
\begin{eqnarray}
\phi= \sqrt{6\alpha}\tanh\left(\frac{\varphi}{\sqrt{6\alpha}M_{pl}}  \right),
\end{eqnarray}
obtaining the following potential,
\begin{eqnarray}\label{alpha}
V(\varphi)=\lambda M_{pl}^4e^{-n\tanh\left(\frac{\varphi}{\sqrt{6\alpha}M_{pl}} \right)},
\end{eqnarray}
where we have introduced the dimensionless parameter $n=\kappa\sqrt{6\alpha}$. Similarly to 
\cite{Guendelman, Guendelman1, Guendelman2} the potential satisfies the {\it cosmological seesaw mechanism}, where the left side of the potential gives a very large energy density -the inflationary side- and the right side gives a very small energy density -the dark energy side. The asymptotic values are $V_{\pm} = \lambda \exp(\pm n)$. The parameter $n$ is the logarithm of the ratios between the energy densities, as $\xi$ in the earlier versions \cite{Guendelman2}. Dealing with this potential at early times, the slow roll parameters are given by
\begin{eqnarray}
\epsilon\equiv \frac{M_{pl}^2}{2}\left( \frac{V_{\varphi}}{V} \right)^2= \frac{n^2}{12\alpha}\frac{1}{\cosh^4\left(\frac{\varphi}{\sqrt{6\alpha}M_{pl}} \right)},
\end{eqnarray}
where we must assume that 
$ \frac{n^2}{12\alpha}>1$ because inflation ends when $\epsilon_{END}=1$, and  the other slow-roll parameter is 
\begin{eqnarray}
\eta\equiv M_{pl}^2\frac{V_{\varphi\varphi}}{V}=\frac{n}{3\alpha}\hspace{-0.1cm}\left[\hspace{-0.1cm}\frac{\tanh\left( \frac{\varphi/M_{pl}}{\sqrt{6\alpha}} \right)}{\cosh^2\left(\frac{\varphi/M_{pl}}{\sqrt{6\alpha}} \right)}
+\frac{n/2}{\cosh^4\left( \frac{\varphi/M_{pl}}{\sqrt{6\alpha}}\right)}\hspace{-0.1cm}\right].\end{eqnarray}

Both slow roll parameters have to be evaluated when the pivot scale leaves the Hubble radius, which will happen for large values of 
$\cosh\left(\frac{\varphi}{\sqrt{6\alpha}M_{pl}} \right)$, obtaining 
\begin{eqnarray}\label{parameters}
\epsilon_*\hspace{-0.1cm}=\hspace{-0.1cm}\frac{n^2}{12\alpha}\frac{1}{\cosh^4\left( \frac{\varphi_*/M_{pl}}{\sqrt{6\alpha}}\right)}, \ \eta_*\hspace{-0.1cm}
\cong\hspace{-0.1cm} -\frac{n}{3\alpha}\frac{1}{\cosh^2\left(\frac{\varphi_*/M_{pl}}{\sqrt{6\alpha}} \right)},
\end{eqnarray}
with $\varphi_*<0$.




Next, we calculate the number of e-folds from the leaving of the pivot scale  to the end of inflation, which for small values of $\alpha$ is given by
\begin{eqnarray}
N=\frac{1}{M_{pl}}\int_{\varphi_*}^{\varphi_{END}}\frac{1}{\sqrt{2\epsilon}}d\varphi
\cong \sqrt{\frac{3\alpha}{4\epsilon_*}},
\end{eqnarray}
so we get the standard form of the spectral index and the tensor/scalar ratio for an $\alpha$-attractor \cite{Linde},
\begin{eqnarray}\label{power}
n_s\cong 
1-6\epsilon_*+2\eta_*\cong
1-\frac{2}{N}, \qquad  r\cong 16\epsilon_*\cong\frac{12\alpha}{N^2}.
\end{eqnarray}



Finally, it is well-known that the power spectrum of scalar perturbations is given by
\begin{eqnarray}
{\mathcal P}_{\zeta}=\frac{H_*^2}{8\pi^2\epsilon_*M_{pl}^2}\sim 2\times 10^{-9}
\end{eqnarray}
and, since in our case $V(\varphi_*)\cong \lambda M^4_{pl}e^{n}$ and, thus,  $H_*^2\cong \frac{\lambda M^2_{pl}}{3}e^{n}$, taking into account that
$\epsilon_*\cong \frac{3\alpha}{16}(1-n_s)^2$ one gets the constraint
\begin{eqnarray}\label{constraint}
\lambda  e^{n}/\alpha\sim 10^{-10},
\end{eqnarray}
where we have chosen as the value of $n_s$ its central value given by the Planck's team, i.e.,  $n_s=0.9649$ \cite{planck18}.



Choosing  for example $\alpha={ 10^{-2}}$, the constraint  (\ref{constraint}) becomes $\lambda e^{n}\sim {10^{-12}}$. On the other hand, at the present time we will have
$\frac{\varphi_0}{\sqrt{6\alpha}M_{pl}}\gg 1$, where $\varphi_0$ denotes the current value of the inflaton field. Hence, we will have $V(\varphi_0)\sim \lambda M_{pl}^4e^{-n}$, which is the dark energy at the present time, meaning that 
\begin{eqnarray}
0.7\cong \Omega_{\varphi, 0}\cong \frac{V(\varphi_0)}{3H^2_0 M_{pl}^2}
\sim \frac{\lambda e^{-n}}{3}\left(\frac{M_{pl}}{H_0}\right)^2.
\end{eqnarray}
Thus, taking for example the value provided by the Planck's team \cite{planck18,planck18a},  $H_0=67.81\; \mbox{Km/sec/Mpc}=5.94\times 10^{-61} M_{pl}$, we get the equations
\begin{eqnarray}
\lambda e^{n}\sim  10^{-12}
\qquad \mbox{and} \qquad \lambda e^{-n}\sim 10^{-120},\end{eqnarray}
whose solution is given by  $n\sim 124$ and $\lambda\sim 10^{-66}$.

\section{Dynamical evolution of the scalar field}
\label{sec-3}

The goal of this section is to   calculate the value of the scalar field and its derivative at the reheating time. 
To do it, first of all we need to calculate the value of the inflaton field and its derivative at the beginning of kination
\cite{Joyce, Spokoiny}, 
which could be calculated as follows:
Taking into account that the slow-roll regime is an attractor, we only need to take initial conditions in the basin of attraction of the slow-roll solution, and thus, integrate the conservation equation up to the moment that the effective EoS parameter was very close to $1$, which is the moment when nearly all the energy density of the scalar field is kinetic.

So, we will take as initial condition the value of the inflaton when the pivot scale leaves the Hubble radius with vanishing temporal derivative (recall that during the slow-roll the kinetic energy is negligible compared with the potential one).
In this way, from the equation (\ref{power}) we get the relation
\begin{eqnarray}
\epsilon_*=\frac{3\alpha}{16}(1-n_s)^2,
\end{eqnarray}
which, together with the expression of $\epsilon_*$ given in (\ref{parameters}), leads to the relation
\begin{eqnarray}
\cosh\left(\frac{\varphi_*}{\sqrt{6\alpha} M_{pl}}  \right)=
\sqrt{\frac{2n}{3\alpha(1-n_s)}},
\end{eqnarray}
whose solution is given by
\begin{eqnarray*}
\varphi_*=\sqrt{6\alpha}M_{pl}\ln\left(\sqrt{\frac{2n}{3\alpha(1-n_s)}}-\sqrt{\frac{2n}{3\alpha(1-n_s)}-1}
\right).
\end{eqnarray*}

Finally, integrating numerically the conservation equation
\begin{eqnarray}
\ddot{\varphi}+3H\dot{\varphi}+V_{\varphi}=0,
\end{eqnarray}
where $H=\frac{1}{\sqrt{3}M_{pl}}\sqrt{\frac{\dot{\varphi}^2}{2}+V(\varphi)}$, and with initial conditions $\varphi_i=\varphi_*$ and $\dot{\varphi}_i=0$, we have obtained for the values $\alpha=10^{-2}$,
$n\cong 124$ and $n_s=0.9649$ the following values at the beginning of the kination period: $\varphi_{kin}\cong  1.1 M_{pl}$ and $\dot{\varphi}_{kin}\cong 6\times 10^{-8} M_{pl}^2$.

When one has these values, analytical calculations can be done disregarding the potential during kination  because in this epoch the potential energy of the field is negligible compared with the kinetic one. Then, since during kination one has $a\propto t^{1/3}\Longrightarrow H=\frac{1}{3t}$, using the Friedmann equation the dynamics in this regime will be obtained solving the equation
\begin{eqnarray}
\frac{\dot{\varphi}^2}{2}=\frac{M_{pl}^2}{3t^2}\Longrightarrow \dot{\varphi}=\sqrt{\frac{2}{3}}\frac{M_{pl}}{t}\Longrightarrow \nonumber\\ 
\varphi(t)=\varphi_{kin}+\sqrt{\frac{2}{3}}M_{pl}\ln \left( \frac{t}{t_{kin}} \right).\end{eqnarray}

\



Thus, at the reheating time, i.e., at the beginning of the radiation era, one has 
\begin{eqnarray}
\varphi_{rh}=\varphi_{kin}+\sqrt{\frac{2}{3}}M_{pl}\ln\left( \frac{H_{kin}}{H_{rh}} \right).
\end{eqnarray}
By  using that at the reheating time (i.e. when the energy density of the scalar field and the one of the relativistic plasma are of the same order) 
the Hubble rate is given by $H_{rh}^2=\frac{2\rho_{rh}}{3M_{pl}^2}$, one gets 
\begin{eqnarray}
\varphi_{rh}=\varphi_{kin}+\sqrt{\frac{2}{3}}M_{pl}\ln\left( \frac{ H_{kin}}{\sqrt{\frac{\pi^2g_{rh}}{45}} \frac{T_{rh}^2}{M_{pl}}}\right)\\
\qquad \mbox{and}  \qquad  { \dot{\varphi}_{rh}=\sqrt{\frac{\pi^2g_{rh}}{15}} T_{rh}^2}\nonumber,\end{eqnarray} 
where we have used that  the energy density and the temperature are related via  the formula
 $\rho_{rh}=\frac{\pi^2}{30}g_{rh}T_{rh}^4$, where the number of degrees of freedom for the Standard Model is $g_{rh}=106.75$  \cite{rg}.

Assuming instant preheating due to the smoothness of the potential \cite{fkl0,fkl,haro19}, we will choose  as the reheating temperature $T_{rh}\cong  10^{9}$ GeV because it is its natural value when this kind of  mechanism 
is the responsible for reheating our Universe.

Then, at the beginning of the radiation era we will have
 \begin{eqnarray}
 \varphi_{rh}\cong  21.5  M_{pl} \qquad \dot{\varphi}_{rh}\cong 1.41 \times 10^{-18} M_{pl}^2.
 \end{eqnarray}

\section{Numerical simulation}\label{sec-4}

\begin{figure*}[t!]
\begin{center}
\includegraphics[width=0.5\textwidth]{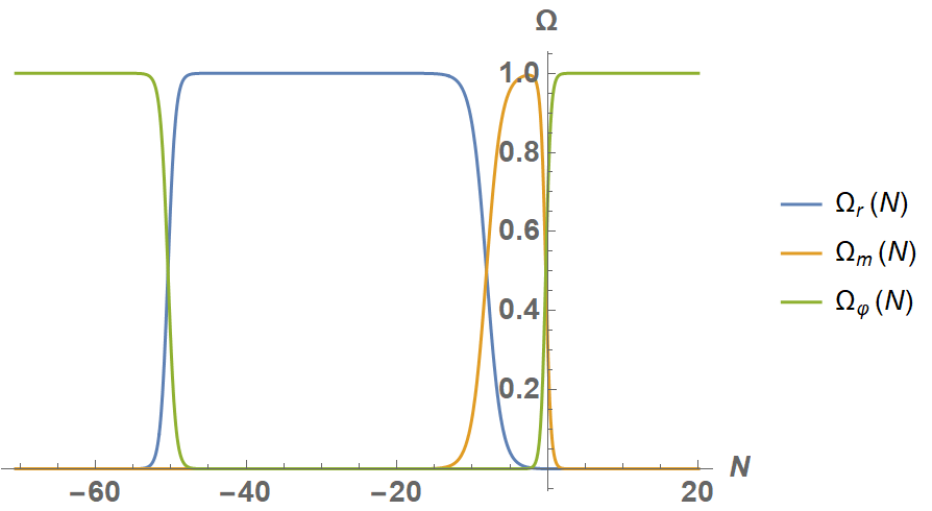}
\includegraphics[width=0.39\textwidth]{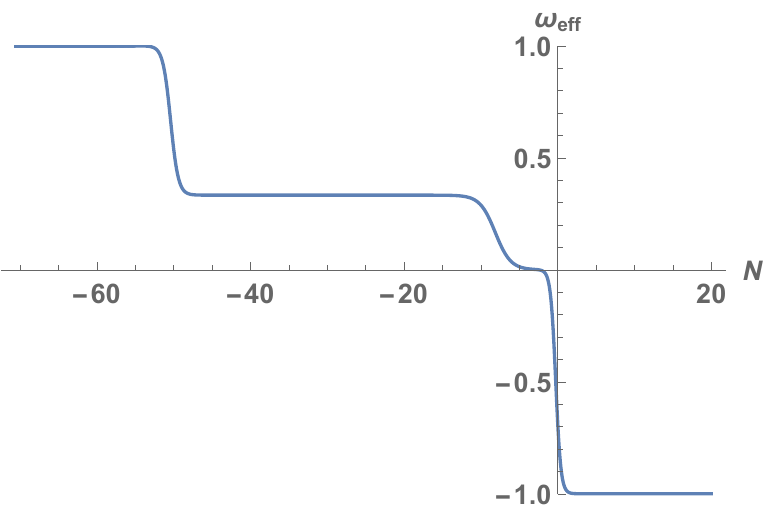}
\end{center}
\caption{\it{\textbf{Left:} The density parameters $\Omega_m=\frac{\rho_m}{3H^2M_{pl}^2}$ (orange curve), $\Omega_r=\frac{\rho_r}{3H^2M_{pl}^2}$ (blue curve) and $\Omega_{\varphi})=\frac{\rho_{\varphi}}{3H^2M_{pl}^2}$, from kination to future times.\\ \textbf{Right:} The effective Equation of State parameter $w_{eff}$, from kination to future times. As one can see in the picture, after kination the Universe enters in a large period of time where radiation dominates. Then, after the matter-radiation equality, the Universe becomes matter-dominated and, finally, near the present, it enters in a new accelerated phase where $w_{eff}$ approaches $-1$.}} \label{fig:Omega}
\end{figure*}
To perform our numerical calculations, first of all we consider the central values obtained in \cite{planck} (see the second column in Table $4$) of  the red-shift at the matter-radiation equality $z_{eq}=3365$, the present value of the ratio of the matter energy density to the critical one $\Omega_{m,0}=0.308$, and, once again, $H_0=67.81\; \mbox{Km/sec/Mpc}=5.94\times 10^{-61} M_{pl}$.
Then, the present value of the matter energy density is $\rho_{m,0}=3H_0^2M_{pl}^2\Omega_{m,0}=3.26\times 10^{-121} M_{pl}^4$, and at matter-radiation equality we will 
have $\rho_{eq}=2\rho_{m,0}(1+z_{eq})^3=2.48\times 10^{-110} M_{pl}^4=8.8\times 10^{-1} \mbox{eV}^4$. So, at the beginning of matter-radiation equality the energy density of the matter  and radiation will be $\rho_{m,eq}=\rho_{r,eq}=\rho_{eq}/2\cong 4.4\times 10^{-1} \mbox{eV}^4$. Therefore, the dynamical equations after the beginning of the radiation can be easily obtained using as a time variable $N\equiv -\ln(1+z)=\ln\left( \frac{a}{a_0}\right)$. Recasting the  energy density of radiation and matter respectively as a function of $N$, we get
\begin{eqnarray}
\hspace{-0.1cm}\rho_{m}(a)={\rho_{m,eq}}\left(\frac{a_{eq}}{a}  \right)^3\rightarrow \rho_{m}(N)= {\rho_{m,eq}}e^{3(N_{eq}-N)} 
\end{eqnarray}
and
\begin{eqnarray}
\rho_{r}(a)={\rho_{r,eq}}\left(\frac{a_{eq}}{a}  \right)^4\rightarrow \rho_{r}(N)={\rho_{r,eq}}e^{4(N_{eq}-N)},
\end{eqnarray}
where  
$N_{eq}\cong -8.121$ denotes the value of the time $N$ at the beginning of the matter-radiation equality. The dynamical system for this scalar field model is obtained 
introducing the  dimensionless variables
 \begin{eqnarray}
 x=\frac{\varphi}{M_{pl}} \qquad \mbox{and} \qquad y=\frac{\dot{\varphi}}{H_0 M_{pl}}.
 \end{eqnarray}
  Thus,  from  the conservation equation $\ddot{\varphi}+3H\dot{\varphi}+V_{\varphi}=0$, one gets   the following   dynamical system,
 \begin{eqnarray}\label{system}
 \left\{ \begin{array}{ccc}
 x^\prime & =& y/\bar H~,\\
 y^\prime &=& -3y-\bar{V}_x/ \bar{H}~,\end{array}\right.
 \end{eqnarray}
 where the prime is the derivative with respect to $N$, $\bar{H}=\frac{H}{H_0}$   and $\bar{V}=\frac{V}{H_0^2M_{pl}^2}$.  Note also that  one can write  
 \begin{eqnarray}
 \bar{H}=\frac{1}{\sqrt{3}}\sqrt{ \frac{y^2}{2}+\bar{V}(x)+ \bar{\rho}_{r}(N)+\bar{\rho}_{m}(N) }~,
 \end{eqnarray}
where we have defined the dimensionless energy densities as
 $\bar{\rho}_{r}=\frac{\rho_{r}}{H_0^2M_{pl}^2}$ and 
 $\bar{\rho}_{m}=\frac{\rho_{m}}{H_0^2M_{pl}^2}$. Finally,  we have to integrate the dynamical system (\ref{system}), with initial conditions $x(N_{rh})=x_{rh}= 21.5$ and $y(N_{rh})=y_{rh}= 2.42\times 10^{42}$ imposing that $\bar{H}(0)=1$, which must be accomplished in order to ensure that the Hubble constant at the present time is the observed one, and where $N_{rh}$ denotes the beginning of reheating, which is obtained imposing that
\begin{eqnarray}
{\rho_{r,eq}}e^{4(N_{eq}-N_{rh})}= \frac{\pi^2}{30}g_{rh} T^4_{rh},\end{eqnarray}
that is,
\begin{eqnarray}
N_{rh}=N_{eq}-\frac{1}{4}\ln\left(\frac{g_{rh}}{g_{eq}}\right)-\ln\left(\frac{T_{rh}}{T_{eq}}\right)\cong -50.68,
\end{eqnarray} 
where we have used that $\rho_{eq,r}=\frac{\pi^2}{30}g_{eq} T^4_{eq}$ with $g_{eq}=3.36$ \cite{rg}, and thus, $T_{eq}\cong 7.81\times 10^{-10}$ GeV. The obtained results are presented in Figure \ref{fig:Omega}, where one can see the similitude with the recent results obtained in \cite{Guendelman2} dealing with Lorentzian Quintessential Inflation.

The Planck's team \cite{planck18} provided the following value of the dark energy EoS parameter at the present time, $w_{de,0}=-1.03\pm 0.03$. So, since the effective EoS parameter is given by
\begin{eqnarray}
w_{eff}=\frac{1}{3}\Omega_r+w_{de}\Omega_{de},
\end{eqnarray}
taking into account that the present value of $\Omega_r$
is approximately $0.0001$ and $\Omega_{de,0}\cong 0.69$, one gets at $1\sigma$ C.L that 
$w_{eff}=-0.712\pm 0.021$, i.e., at $2\sigma$ C.L we have
$-0.754\leq w_{eff,0}\leq -0.67$, which is compatible with our model, as one can see on the right-hand side of  Figure \ref{fig:Omega}. In fact, in our case we have obtained $w_{eff,0}\cong-0.68$.


\section{Observational Constraints} 
\label{sec:obs}
Next we describe the observational data sets along with the relevant statistics in constraining the model. The data set incorporates few different measurements. 

\subsubsection{Direct measurements of the Hubble expansion}
\textbf{Cosmic Chronometers (CC)}: The data set exploits the evolution of differential ages of passive galaxies at different redshifts to directly constrain the Hubble parameter \cite{Jimenez:2001gg}. We use uncorrelated 30 CC measurements of $H(z)$ discussed in \cite{Moresco:2012by,Moresco:2012jh,Moresco:2015cya,Moresco:2016mzx}. Here, the corresponding $\chi^2_{H}$ function reads
\begin{equation}
\chi^{2}_{H} = \sum_{i=1}^{30}\left(\frac{H_{i} - H_{pred}(z_i)}{\Delta H_i}\right)^2,
\end{equation}
where $H_{i}$ is the observed Hubble rates at redshift $z_{i}$ ($i=1,...,N$) and $H_{pred}$ is the predicted one from the model.

\subsubsection{Standard Candles}
As \textbf{Standard Candles (SC)} we use measurements of the Pantheon Type Ia supernova (SnIa) \cite{Scolnic:2017caz}. The model parameters of the models are to be fitted with by comparing the observed $\mu _{i}^{obs}$ value to the theoretical $\mu _{i}^{th}$ value of the distance moduli which are the
logarithms
\begin{equation}
 \mu=m-M=5\log _{10}(D_{L})+\mu _{0},   
\end{equation}
where $m$ and $M$ are the apparent and absolute magnitudes and $\mu_{0}=5\log \left( H_{0}^{-1}/Mpc\right) +25$ is the nuisance parameter that
has been marginalized. The luminosity distance is defined by
\begin{eqnarray}
D_L(z) &=&\frac{c}{H_{0}}(1+z)\int_{0}^{z}\frac{dz^{\ast }}{%
E(z^{\ast })}.
\end{eqnarray}%
Here $\Omega _{k}=0$ (flat spacetime). Following standard lines, the chi-square function of the standard candles is given by
\begin{equation}
\chi^{2}_{\text{SC}}\left(\phi^{\nu}_{\text{s}}\right)={ \bf \mu}_{\text{s}}\,{\bf C}_{\text{s},\text{cov}}^{-1}\,{\bf \mu}_{\text{s}}^{T}\,,
\end{equation}
where ${\bf \mu}_{\text{s}}=\{\mu_{1}-\mu_{\text{th}}(z_{1},\phi^{\nu})\,,\,...\,,\,\mu_{N}-\mu_{\text{th}}(z_{N},\phi^{\nu})\}$ and the subscript $\text{`s'}$ denotes SnIa and QSOs. For the SnIa data the covariance matrix is not diagonal and the distance modulus is given as $\mu_{i} = \mu_{B,i}-\mathcal{M}$, where $\mu_{B,i}$ is the maximum apparent magnitude in the rest frame for redshift $z_{i}$ and $\mathcal{M}$ is treated as a universal free parameter \cite{Scolnic:2017caz}, quantifying various observational uncertainties. It is apparent that $\mathcal{M}$ and $h$ parameters are intrinsically degenerate in the context of the Pantheon data set, so we can not extract any information regarding $H_{0}$ from SnIa data alone. 
\begin{figure*}[t!]
 	\centering
\includegraphics[width=0.9\textwidth]{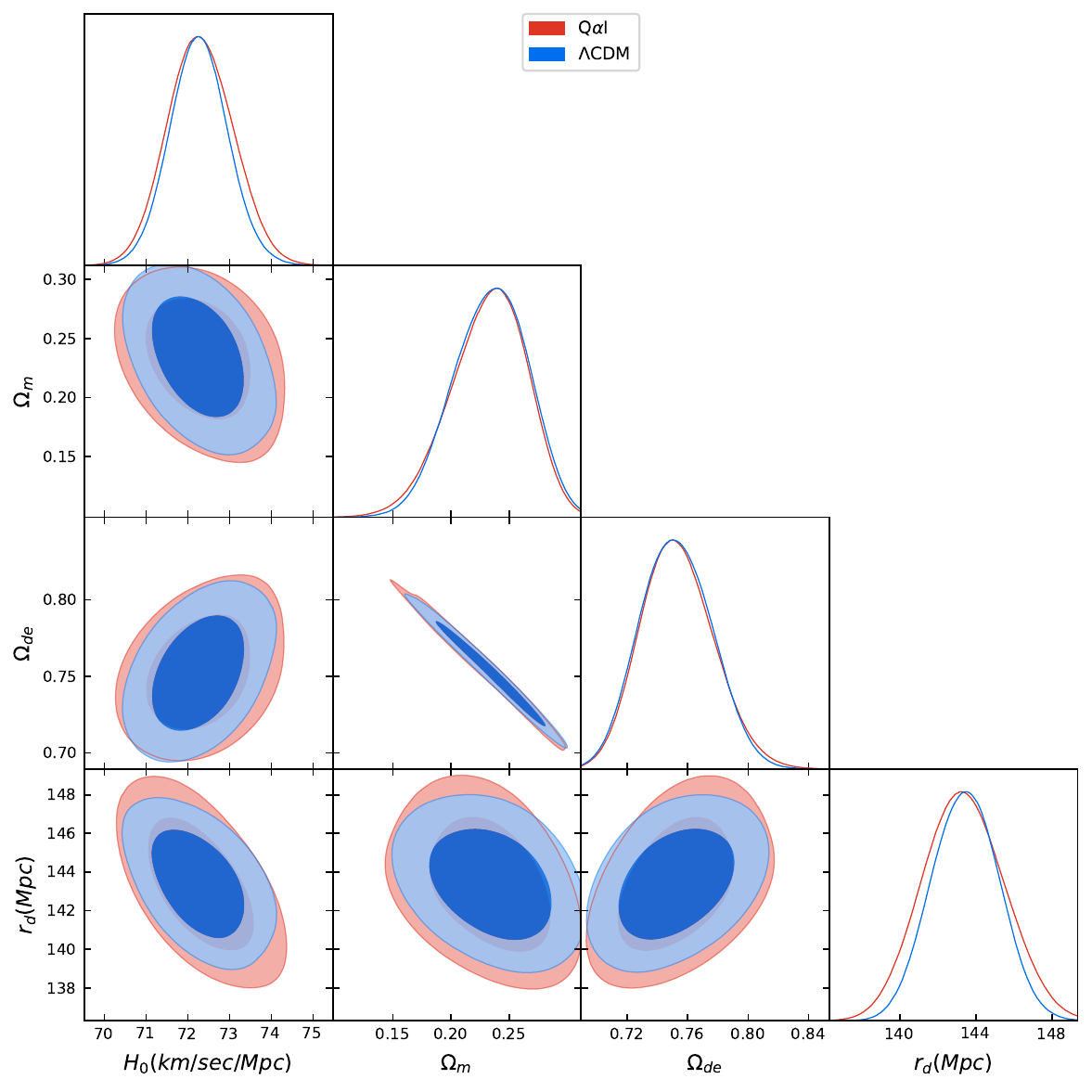}
\caption{\it{The posterior distribution for different measurements with the {Quintessential $\alpha$-attractor Inflation (Q$\alpha$I)} model with $1 \sigma$ and $2\sigma$ for $\Omega_m$, $\Omega_{de}$, $H_0$ and $r_d$.}}
 	\label{fig:corPlot}
\end{figure*}
\begin{table*}[t!]
\tabcolsep 5.5pt
\vspace{1mm}
\centering
\begin{tabular}{ccc} \hline \hline
Parameter & {Q$\alpha$I } & $\Lambda$CDM \\ \hline \hline
$H_0 (km/sec/Mpc)$ & $72.25 \pm0.74$ & $72.24 \pm 0.65$ \\
$\varphi_0/M_{pl}$ & $22.46 \pm1.419$ & - \\
$\dot{\varphi}_0/(H_0  M_{pl}) \, 10^{-71}  $ & $5.09 \pm2.858$ & - \\
$\Omega_m$ & $0.2323 \pm0.0286$ &$0.2393 \pm0.02751$ \\
$\Omega_{de}$  & $0.7535 \pm0.02092$ & $ 0.7489 \pm0.2027$ \\
$n$ & $122.1 \pm2.021$ & - \\
$\alpha$ & $0.2760 \pm0.1448$ & - \\
$r_d (Mpc)$ &$143.4 \pm1.941$& $143.5 \pm1.577$\\
$B_{ij}$ &$-80.54$& $-79.43$\\
\hline\hline
\end{tabular}
\caption[]{\it{The best fit values for the discussed model. The values $\varphi_0$, $\dot{\varphi}_0$ denote the current values of the scalar field and its derivative.}}
\label{tab:Res}
\end{table*}

\subsubsection{Baryon acoustic oscillations}
We use uncorrelated data points from different \textbf{Baryon Acoustic Oscillations (BAO)}. BAO are a direct consequence of the strong coupling between photons and baryons in the pre-recombination epoch. After the decoupling of photons, the over densities in the baryon fluid evolved and attracted more matter, leaving an imprint in the two-point correlation function of matter fluctuations with a characteristic scale of around $r_d \approx 147$ Mpc that can be used as a standard ruler and to constrain cosmological models. Studies of the BAO feature in the transverse direction provide a measurement of $D_H(z)/r_d = c/H(z)r_d$, with the comoving angular diameter distance being \cite{Hogg:2020ktc,Martinelli:2020hud}
\begin{equation}
D_M= \int_0^z\frac{c \, dz'}{H(z')}.
\end{equation}
The angular diameter distance $D_A=D_M/(1+z)$ and $D_V(z)/r_d$ are a combination of the BAO peak coordinates above, namely
\begin{equation}
    D_V(z) \equiv [ z D_H(z) D_M^2(z) ]^{1/3}.
\end{equation}
The surveys provide the values of the measurements at some effective redshift. We employ the following BAO data points, collected in \cite{Benisty:2020otr} from \cite{Percival:2009xn,Beutler:2011hx,Busca:2012bu,Anderson:2012sa,Seo:2012xy,Ross:2014qpa,Tojeiro:2014eea,Bautista:2017wwp,deCarvalho:2017xye,Ata:2017dya,Abbott:2017wcz,Molavi:2019mlh}, in the redshit range $0.106 < z < 2.34$. Since \cite{Benisty:2020otr} proves the un-correlation of this data set, 
\begin{equation}
\chi^2_{BAO} = \sum_{i=1}^{17}\left(\frac{D_{i} - D_{pred}(z_i)}{\Delta D_i}\right)^2,
\end{equation}
where $D_{i}$ is the observed distant module rates at redshift $z_{i}$ ($i=1,...,N$) and $D_{pred}$ is the predicted one from the model.

\subsubsection{Cosmic Microwave Babkground}

Finally we take the \textbf{CMB Distant Prior} measurements \cite{Chen:2018dbv}. The distance priors provide effective information of the CMB power spectrum in two aspects: the acoustic scale $l_A $
characterizes the CMB temperature power spectrum in the transverse direction, leading to the variation of the peak
spacing, and the “shift parameter” $R$ influences the CMB temperature spectrum along the line-of-sight direction,
affecting the heights of the peaks, which are defined as follows:
\begin{equation}
\begin{split}
l_A =  (1 + z_d) \frac{\pi D_A(z_d)}{r_d}, \\ R(z_d) = \frac{\sqrt{\Omega_m} H_0}{c}(1 + z_d)  D_A(z_d),  
\end{split}
\end{equation}
with its corresponding covariance matrix (see Table I in \cite{Chen:2018dbv}). The BAO scale is set by the sound horizon at the drag epoch $z_d \approx 1060$ when photons and baryons decouple, given by
\begin{equation}
r_d = \int_{z_d}^{\infty} \frac{c_s(z)}{H(z)} dz,
\end{equation}
where $c_s \approx c \left(3 + 9\rho_b /(4\rho_\gamma) \right)^{-0.5}$ is the speed of sound in the baryon-photon fluid with the baryon and photon densities being $\rho_b(z)$ and $\rho_\gamma(z)$ respectively \cite{Aubourg:2014yra}. However, in our analysis we used $r_d$ as an independent parameter. The $\chi^2_{CMB}$ is defined in \cite{Chen:2018dbv}. 

We include the latest measurement of the Hubble parameter:
\begin{equation}
H_0 = (73.2 \pm 1.3) \text{km/s/Mpc}
\end{equation}
reported by \cite{Riess:2020fzl}. The measurement presents an expanded sample of 75 Milky Way Cepheids with Hubble Space Telescope (HST) photometry and Gaia EDR3 parallaxes which uses the extragalactic distance ladder in order to recalibrate and refine the determination of the Hubble constant.  {The combination is related via the relation}
\begin{equation}
\chi^2_{Hub} = \left(\frac{H_0 - 73.2}{1.3}\right)^2.
\end{equation}
 {The $\chi^2_{Hub}$ estimates the deviation from the latest measurement of the Hubble constant.}
\subsubsection{Joint analysis and model selection}
In order to perform a joint statistical analysis of $4$ cosmological probes we need to use the total likelihood function, consequently the  $\chi^2_{\text{tot}}$ expression is given by
\begin{equation}
\chi_{\text{tot}}^2 = \chi_{CMB}^2 + \chi_{H}^2 + \chi_{SC}^2 + \chi_{BAO}^2 + \chi^2_{Hub}.
\end{equation}
Regarding the problem of likelihood maximization, we use an affine-invariant Markov Chain Monte Carlo sampler \cite{ForemanMackey:2012ig}, as it is implemented within the open-source packaged $Polychord$ \cite{Handley:2015fda} with the $GetDist$ package \cite{Lewis:2019xzd} to present the results. The prior we choose is with a uniform distribution, where $\Omega_{m} \in [0.;1.]$, $\Omega_{de}\in[0.;1 - \Omega_{m}]$, $\Omega_{r}\in[0.;1 - \Omega_{m} - \Omega_{de}]$, $H_0\in [50;100]$Km/sec/Mpc, $r_d\in [130;160]$Mpc. For the scalar field final condition we imposed $\phi \in [20;25] $.

 {Furthermore, we use the logarithmic Bayes factor defined as}
\begin{equation}
\log(B_{01})=\log(Z_0)-\log(Z_1),    
\end{equation}
 {where $Z_i$ is the logarithmic marginalised evidence reported by Polychord  \cite{doi:10.1080/01621459.1995.10476572}. For the logarithmic Bayes factor, a difference of $\log(B_{01})\in [1/2,1]$ is substantial in favor of $Z_0$, $\log(B_{01})\in [1,2]$ is strong and $\log(B_{01})>2$ is decisive. In the case of negative values, the same applies for $Z_1$. }

\subsubsection{Results}
\begin{figure*}[t!]
\begin{center}
\includegraphics[width=0.9\textwidth]{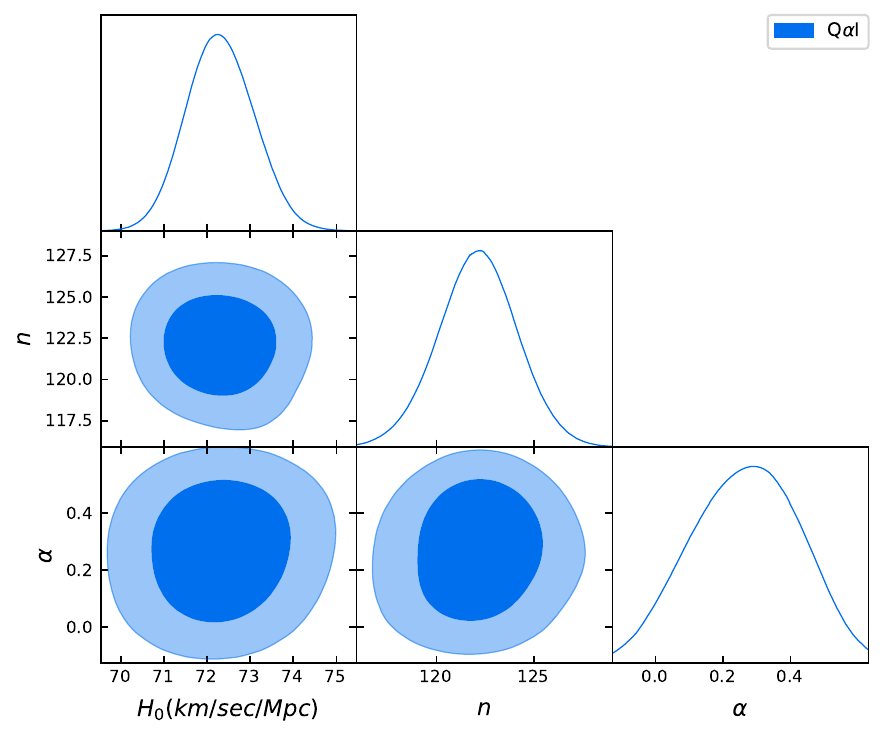}
\end{center}
\caption{\it{The posterior distribution for the Q$\alpha$I model with $1 \sigma$ and $2\sigma$ C.L., for the Hubble parameter vs. the parameter $n$ and $\alpha$. The data set include Baryon Acoustic Oscillations dataset, Cosmic Chronometers, the Hubble Diagram from Type Ia supernova and the CMB}.\label{fig:fitH0}}
\end{figure*}

 {Figure \ref{fig:corPlot} shows the posterior distribution of the data fit with the best fit values at Table \ref{tab:Res}. The posterior one for the additional parameters is described in Fig \ref{fig:fitH0}. The Q$\alpha$I model is a viable model and can describe early times as well as late times. Actually there is no distinguishable difference between the $\Lambda$CDM fit and the Q$\alpha$I, since the potential in that regime includes a slow roll behavior. In conclusion, this model includes the inflationary period and predicts the large difference between the inflationary and the late dark energy, while the standard models do not predict that. But also for the measurements from the observed Universe there is no distinguishable difference between the standard models and the Q$\alpha$I model. }

 {In order to complete our analysis, we use the Bayesian Evidence. The difference between the models yields $\Delta B_{ij}  = 1.11$, which implies a slight preference for the $\Lambda$CDM model. It seems that the difference is due to the additional parameters that the model suggests. However, as we said, this additional parameter gives a natural explanation for the DE differences. But statistically there is a slight preference for the $\Lambda$CDM model.}

\begin{figure*}[t!]
 	\centering
\includegraphics[width=0.44\textwidth]{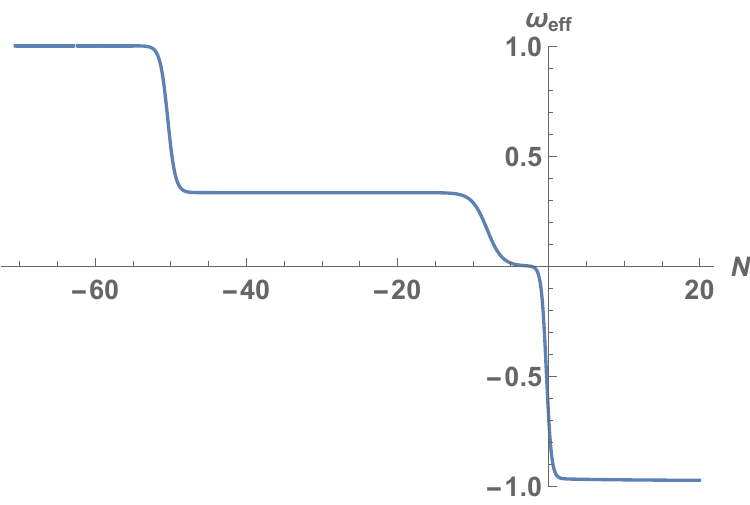}
\includegraphics[width=0.44\textwidth]{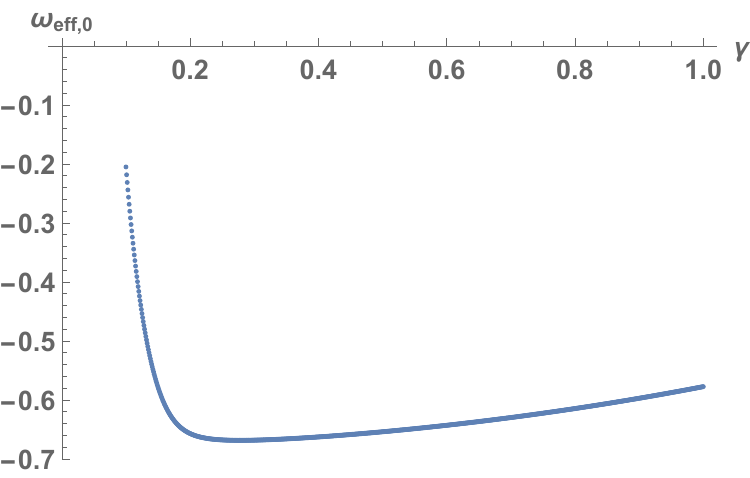}
\caption{\it{\textbf{Left:} {Evolution of the Equation of State parameter $w_{eff}$ from the kination phase to late times for $\gamma=0.277$ according to the system \eqref{dynamical}, taking as initial conditions the ones obtained from the numerical simulation carried out in Section \ref{sec-4}. 
We see that $w_{eff}$ is $1$ during the kination phase that takes place in quintessential inflation models after inflation. Then reheating occurs and $w_{eff}$ becomes $1/3$, which is maintained during all the radiation phase until the matter-radiation equality. And then it finally effectively converges to $\gamma^2/3-1$ obtaining an eternal accelerating Universe.}\\
\textbf{Right:} Equation of State parameter at the present time for different set of values of $0<\gamma<\sqrt{2}$. All of them lie outside of the $2\sigma$ C.L. Planck's observational data, including $\gamma=0.277$ with $w_{eff,0}=-0.669$, though this value lies very close to it.
}} 
 	\label{fig:weffgamma}
\end{figure*} 

\section{$\alpha$-attractors WITH A FINE-TUNED COSMOLOGICAL CONSTANT}
\label{sec-5}

In this section we review  the treatment of $\alpha$-attractor done in \cite{Dimopoulos:2017zvq}. First of all, one has to introduce a Cosmological Constant (CC) with the following form,
$\Lambda=\lambda M_{pl}^2e^{-n}$, and thus, adding to the Lagrangian the term $\Lambda M_{pl}^2$ leads to the following effective potential,
\begin{eqnarray}\label{alpha1}
V(\varphi)=\lambda M_{pl}^4e^{-n}\left(e^{n\left(1-\tanh\left(\frac{\varphi}{\sqrt{6\alpha}M_{pl}} \right)\right)} -1\right).
\end{eqnarray}

During inflation $\varphi<0$ and the potential becomes as (\ref{alpha}) because $e^{-n}\ll 1$. So, as we have shown for the potential (\ref{alpha}), for small values of $\alpha$ inflation works also well when this CC is introduced in the model.

In the same way, for large values of the scalar field the potential will become
\begin{eqnarray}
V(\varphi)=2n\lambda e^{-n}M_{pl}^4e^{-\gamma\varphi/M_{pl}},
\end{eqnarray}
with $\gamma\equiv \sqrt{\frac{2}{3\alpha}}$. It is well-known \cite{barreiro, liddle} that for an exponential potential a late time eternal acceleration 
is  achieved when $\gamma<\sqrt{2}$, that is, for  $\alpha>1/3$.  Effectively, as has been shown in \cite{hap19}, by
 introducing the dimensionless variables 
\begin{eqnarray}
\tilde{x}\equiv \frac{\dot{\varphi}}{\sqrt{6}M_{pl}H} \quad \mbox{and} \quad \tilde{y}\equiv \frac{\sqrt{V}}{\sqrt{3}M_{pl}H},
\end{eqnarray}
after the matter-radiation equality the dynamical system (\ref{system}) can be written as follows,
 \begin{eqnarray}\label{dynamical}
 \left\{\begin{array}{ccc}
 \tilde{x}'&=& -3\tilde{x}+\sqrt{\frac{3}{2}}\gamma \tilde{y}^2+\frac{3}{2}\tilde{x}\left[\tilde{x}^2-\tilde{y}^2+1\right]\\
  \tilde{y}'&=& -\sqrt{\frac{3}{2}}\gamma\tilde{x}\tilde{y}+\frac{3}{2}\tilde{y}\left[\tilde{x}^2-\tilde{y}^2+1\right]\ \end{array},
 \right.
 \end{eqnarray} 
together with the constraint
\begin{eqnarray}
\tilde{x}^2+\tilde{y}^2+\Omega_m=1.
\end{eqnarray}
The system (\ref{dynamical}) has the following attractor fixed point, $\tilde{x}=\frac{\gamma}{\sqrt{6}}$ and $\tilde{y}=\sqrt{1-\frac{\gamma^2}{6}}$, 
which depicts an attractor (tracker) solution with $w_{eff}=\tilde{x}^2-\tilde{y}^2=\frac{\gamma^2}{3}-1$ and $\Omega_{\varphi}=1$. For this reason, if one demands an accelerated period at late times one has to choose $\gamma^2<2$, as seen in Figure \ref{fig:weffgamma} for the case $\gamma=0.277$. In addition, as has been shown in \cite{hap19}, this tracker solution is given by
\begin{eqnarray}
\varphi_{tra}(N)=-\frac{n}{\gamma}M_{pl}+NM_{pl}+\frac{M_{pl}}{\gamma}\ln\left(\frac{4n\lambda}{(6-\gamma)^2} \right)\\
+\frac{2M_{pl}}{\gamma}\ln\left(\frac{M_{pl}}{H_0}\right)\nonumber.
\end{eqnarray}

However, since in this case one has to choose $\alpha>1/3$, the calculation of the spectral index and the ratio of tensor to scalar perturbations changes a little bit with respect to the case $\alpha\ll 1$, obtaining (see for details \cite{Dimopoulos:2017zvq}):
\begin{eqnarray}
n_s\cong 1-\frac{2}{N+\frac{\sqrt{3\alpha}}{2}} , \quad r=\frac{12\alpha}{\left(N+\frac{\sqrt{3\alpha}}{2}\right)^2}.\end{eqnarray}
To end with the case $\gamma\leq \sqrt{2}$, we have numerically checked that, in order to obtain at the present time an effective EoS parameter compatible with the Planck's data, there is a very narrow range of values of $\alpha$, as is shown in Figure \ref{fig:weffgamma} . In fact, the only value which might be considered viable is
$\gamma \cong  0.277 \Longleftrightarrow \alpha\cong 8.688$, which lies very close to the lower bound of the $2\sigma$ CL of the allowed values. So, this is not at all sufficient to prove or disprove the viability of the CC model.

On the other hand, in the case $\gamma>\sqrt{3}\Longrightarrow \alpha<2/9$, the dynamical system (\ref{dynamical}) has another  fixed point, namely 
$\tilde{x}=\tilde{y}= \sqrt{\frac{3}{2}}\frac{1}{\gamma}$, which corresponds to a matter-dominated Universe because $w_{eff}=0$. In that case, it is argued in \cite{Dimopoulos:2017zvq}
that, when $\sqrt{3}<\gamma<2\sqrt{6}\Longrightarrow 1/36<\alpha<2/9$, the scalar field may dominate for a brief period obtaining a short period of acceleration. However, we have not been able to find the numerical values of the parameters $\lambda$ and $n$ satisfying the constraint $\lambda e^n/\alpha\sim 10^{-10}$ provided by the power spectrum of scalar perturbations and the essential identity $\bar{H}(0)=1$ at the present time. That is, when one adds this cosmological constant, for values of $\alpha$ less than $2/9$, from our viewpoint it is impossible to unify the early and late time acceleration of our Universe.

\section{Power law potentials}
In the Lagrangian (\ref{lagrangian}) one can replace the exponential potential by a power law with the form
\begin{eqnarray}
V_s(\phi)=\lambda M_{pl}^4\left(1-e^{-\beta}\frac{\phi}{\sqrt{6\alpha}}   \right)^s,
\end{eqnarray}
where $\lambda$, $\beta$ and $\alpha$ are positive dimensionless variables and $s$ an odd number. We also assume that
$\beta$ is close to zero. Then, in terms of the scalar field $\varphi$, the potential becomes
\begin{eqnarray}
V_s(\varphi)=\lambda M_{pl}^4\left(1-e^{-\beta}\tanh\left(\frac{\varphi}{\sqrt{6\alpha}M_{pl}} \right)   \right)^s,
\end{eqnarray}
which, for small values of $\alpha$, belongs to the class of $\alpha$-attractors (see for instance \cite{kallosh4}).

To obtain the value of the parameters $\lambda$ and $\beta$, we follow the same method as in Section \ref{sec-2}, getting
\begin{eqnarray}
2^s\lambda/\alpha\sim 10^{-10}, \qquad \lambda (1-e^{-\beta})^s\sim 10^{-120}.
\end{eqnarray}

Choosing for example $\alpha\sim 10^{-2}$, we get
\begin{eqnarray}
\lambda\sim 10^{-12}2^{-s}\quad \mbox{and}
\quad \beta\sim -\ln(1-2\times 10^{-108/s}), 
\end{eqnarray}
which, for  $s=1$, leads to
\begin{eqnarray}
\lambda\sim 5\times 10^{-13}\quad \mbox{and}
\quad \beta\sim 2\times 10^{-108}, 
\end{eqnarray}
 for $s=3$ to
\begin{eqnarray}
\lambda\sim 1.25\times 10^{-13}\quad \mbox{and}
\quad \beta\sim 2\times 10^{-36},
\end{eqnarray}
and so on.

Then, since for small values of $s$ one has $\beta\cong 0$, we can safely assume that $\beta=0$, and the potential becomes $V_s(\varphi)=\lambda M_{pl}^4\left(1-\tanh\left(\frac{\varphi}{\sqrt{6\alpha}M_{pl}} \right)   \right)^s$, which for large values of the scalar field (which happens at the present time)  has the following exponential form,
\begin{eqnarray}
V_s(\varphi)\cong 2^s\lambda M_{pl}^4 e^{-\frac{2s\varphi}{\sqrt{6\alpha}M_{pl}}}
\cong \alpha 10^{-10} M_{pl}^4 e^{-\gamma\varphi/M_{pl}},
\end{eqnarray}
where now $\gamma=\sqrt{\frac{2}{3\alpha}}s$. Thus, as we have already commented in Section \ref{sec-5}, in order to have an accelerated expansion at late times, the value of $\gamma$ must be smaller than $\sqrt{2}$, that is, one has to choose $\alpha>s^2/3$. In fact, to agree with the Planck's observational data, one has to choose the parameter $\alpha$ to satisfy, at the present time,
$w_{eff,0}=-0.712\pm 0.03$.

Unfortunately, in these power law models inflation never ends. Effectively, choosing for simplicity $s=1$, the main slow-roll parameter is given by 
\begin{eqnarray}
\epsilon=\frac{1}{12\alpha}
\frac{1}{\cosh^4\left( \frac{\varphi}{\sqrt{6\alpha}M_{pl}} \right)}
\frac{1}{\left(1-\tanh\left( \frac{\varphi}{\sqrt{6\alpha}M_{pl}}
\right)\right)^2}\nonumber\\
=\frac{1}{12\alpha}
\left(1+\tanh\left( \frac{\varphi}{\sqrt{6\alpha}M_{pl}}
\right)\right)^2.
\end{eqnarray}
Thus, at the end of inflation ($\epsilon=1$),
we will have 
\begin{eqnarray}
\tanh\left( \frac{\varphi_{END}}{\sqrt{6\alpha}M_{pl}}
\right)=-1+\sqrt{12\alpha},
\end{eqnarray}
which does not have solution for $\alpha>1/3$. Therefore, we can conclude that these power law potentials must be disregarded to have a QI behavior.

\section{Concluding remarks}
\label{sec-summary}
We have studied different Quintessential Inflation potentials such as exponential or power law potentials in the context of $\alpha$-attractors. We have shown that the potential which provides the best results compatible with the current observational data is the exponential one without any kind of Cosmological Constant. In fact, the behavior of the dynamics provided by an exponential $\alpha$-attractor potential is very similar to the dynamics in  the so-called Lorentzian Quintessential Inflation, where at very late times the effective EoS parameter converges to $-1$.

 {We have also verified that this model fits statistically very well with the observational data sets coming from Type Ia Supernova, Cosmic Microwave Background and the Cosmic Chronometers. The {Q$\alpha$I } model is a viable model from the data fit. There is a slight preference for the $\Lambda$CDM model from the Bayesian Evidence. However, this model still unifies naturally inflationary and late time dark energy behavior and explains the difference between the energy density values from these epochs.}

On the contrary, the introduction of a Cosmological Constant leads for values of the parameter $\alpha$ greater than $1/3$, at the present time,  to an effective EoS parameter that does not enter at the $2\sigma$ C.L, in the region provided by the Planck's team, and even worse, for $\alpha<2/9$ the model is unable to depict both the early and late time acceleration of the Universe. In addition, for power law potentials, in the context of $\alpha$-attractors, the inflationary regime never finishes, which invalidates its viability.

\acknowledgments
D. Benisty and E. I. Guendelman thank Ben Gurion University of the Negev for a great support. D. Benisty also thanks to the Grants Committee of the Rothschild and the Blavatnik Fellowships for generous supports. The investigation { of J. de Haro} has been supported by MINECO (Spain) grant  MTM2017-84214-C2-1-P and  in part by the Catalan Government 2017-SGR-247. This work has been supported by the European COST actions CA15117 and CA18108.

\bibliographystyle{apsrev4-1}
\bibliography{ref}

\end{document}